\documentclass[pdflatex,sn-nature]{sn-jnl}

\usepackage{graphicx}
\usepackage{amsmath,amssymb}
\usepackage{booktabs}
\usepackage{xcolor}
\usepackage{enumitem}
\usepackage{array}
\usepackage{url}
\newcolumntype{L}[1]{>{\raggedright\arraybackslash}p{#1}}

\newcommand{\figdir}{output}

\begin{document}

\title[Divergent energy changes from AI adoption]{AI adoption induces divergent net energy changes across economic sectors}

\author*[1]{\fnm{Wei} \sur{He}}\email{wei.4.he@kcl.ac.uk}

\author[2]{\fnm{Daoping} \sur{Wang}}

\author[3]{\fnm{Hanqi} \sur{Yan}}

\author[4]{\fnm{Yang} \sur{Wang}}

\author[5]{\fnm{Sai} \sur{Gu}}

\affil*[1]{\orgdiv{Department of Engineering}, \orgname{King's College London}, \orgaddress{\city{London}, \country{United Kingdom}}}

\affil[2]{\orgdiv{Department of Geography}, \orgname{King's College London}, \orgaddress{\city{London}, \country{United Kingdom}}}

\affil[3]{\orgdiv{Department of Informatics}, \orgname{King's College London}, \orgaddress{\city{London}, \country{United Kingdom}}}

\affil[4]{\orgname{Energy System Catapult}, \orgaddress{\city{Birmingham}, \country{United Kingdom}}}

\affil[5]{\orgdiv{School of Engineering}, \orgname{University of Warwick}, \orgaddress{\city{Coventry}, \country{United Kingdom}}}

\abstract{Energy planning for artificial intelligence focuses on data-centre electricity, missing the induced operational energy change caused by the deployment of AI in commercial buildings, factories and freight networks. Here we map occupation-level AI exposure onto sector energy use and apply a Monte Carlo (MC) joint supply--demand decomposition to estimate each sector's net energy change. Our results show that the US adoption-side energy envelope---the operational energy exposed to AI---is 12.1~Q theoretical and $\sim$1.4~Q observed (1~Q $\approx$293~TWh, summed across electricity, gas, petroleum and process fuels); this measures the scope of exposed energy, not consumption. Decomposing this envelope at full adoption reveals divergent sector net signs: Commercial saves 0.22~Q while Industrial ($+$1.25~Q) and Transport ($+$1.12~Q) increase, each sign robust across 88--99\% of parameter draws. The induced net change aggregates to $+$2.16~Q (90\% MC range [+0.52, +4.12]; $+$1.1~Q under a conservative price-channel conversion of the rebound anchors)---several times the $\sim$0.6~Q of current US data-centre electricity that AI energy planning targets. These net changes vary geographically when projected onto each state's occupational and energy end-use mix. Industrial- and freight-heavy states (Texas, Louisiana, Indiana) primarily carry the increase, while commercial-dominated states (New York, Massachusetts, DC) see substantially smaller net changes. We also transfer the analysis to the UK and show an energy envelope of 1.9~Q out of a 3.7~Q national total. Therefore, adoption-side energy is the larger, geographically variable component of AI's footprint, requiring end-use energy surveys to track AI deployment and the resulting task and occupational shifts alongside compute-side forecasting.}


\maketitle

% ═════════════════════════════════════════════════════════════════════════════
% SECTION 1: Introduction (untitled per Nature Energy style)
% Claims: C-009 (planning gap), C-010 (capability), C-001 (preview)
% ═════════════════════════════════════════════════════════════════════════════
\section{Introduction}

% P1: DC supply side, affirm-first (C-009 partial)
Data-centre electricity has emerged as a primary axis of grid-system planning. US data centres consumed 176~TWh in 2023 and could reach 325--580~TWh by 2028\cite{shehabi2024dc,eia_demand_2026,doe_datacenter_pathways_2024}; globally, 415~TWh in 2024 (around 1.5\% of electricity) is projected to reach 945~TWh by 2030\cite{iea_energy_ai_2025}. Grid operators, utilities and regulators have moved compute load to the centre of capacity planning, connection queuing and infrastructure investment, with policy responses ranging from Ireland's grid-connection moratorium for new data centres to record PJM capacity-auction prices and the US Department of Energy's data-centre pathways report\cite{doe_datacenter_pathways_2024}. All these instruments target one well-defined problem: electricity scarcity at identifiable data-centre point loads, met with near-term grid capacity build-out.

% P2: Coverage asymmetry — what the planning frame measures vs what it does not (C-009)
However attentive, this monitoring covers only where AI is trained and served, not where AI-mediated work is delivered. AI is already entering scheduling, process control, dispatch and administrative workflows in commercial buildings, industrial facilities and freight networks\cite{anthropic_labor_2026}. These sectors consume electricity alongside natural gas, petroleum and process fuels; on the US end-use convention of quadrillion British thermal units (Q; $1$~Q $\approx 293$~TWh, summed on an equal energy-content basis across these vectors), they totalled 35.9~Q\cite{eia_cbecs_2018,eia_mecs_2018,bts_nts_2021}, an order of magnitude above US data-centre electricity in 2023 at 0.6~Q (176~TWh). Yet existing energy-and-AI assessments treat adoption-side effects only qualitatively\cite{iea_energy_ai_2025} or alongside compute-side pathways without sectoral resolution\cite{doe_datacenter_pathways_2024}. The result is an asymmetric planning gap: compute load is tracked in detail while the much larger operational energy AI reshapes goes unmeasured.

% P3: Why the gap is active now — capability evidence as load-bearing answer (C-010)
AI deployment is evolving rapidly because frontier capability has moved from chatbot prompts to multi-step operational tasks. Over three years the time horizon at which leading AI agents succeed half the time has expanded from minutes to hours\cite{metr_horizon_2026}, and published experiments across software engineering, professional writing, customer service, legal review and clinical diagnostics report 10--55\% completion-time reductions\cite{peng2023impact,noy2023experimental,brynjolfsson2023generative} (SI~1). Wide deployment remains early, with success rates much lower for tasks requiring physical manipulation or on-site coordination\cite{metr_limitations_2026}; the trajectory, however, points to operational integration rather than isolated assistance. Information-adjacent roles within energy-consuming sectors are already adopting, with measurable uptake extending beyond commercial knowledge work into industrial and transport operations\cite{anthropic_labor_2026}. As these capabilities enter the workflows of commercial buildings, factories and freight networks, they begin to interact with the energy those workflows already consume. Without balanced measurement, this gap tilts infrastructure investment toward compute and can throttle AI's diffusion into the operational sectors where its value is largest.

% P4: "Here we..." — what we do, method framed as enabling tool (C-001, C-014)
Here we quantify both the adoption-side energy exposure envelope and the net energy change it implies. We map occupation-level AI exposure onto US end-use energy surveys\cite{eia_cbecs_2018,eia_mecs_2018,bts_nts_2021} under two complementary attributions. Approach~A scales sector energy directly by employment-weighted exposure (top-down, economics-proportional). Approach~B weights each occupation by its energy relevance (bottom-up, occupation-level). Together they keep the framework explainable at both macro and micro levels. To trace efficiency and rebound through sectoral energy use, we then apply a joint supply--demand decomposition. Task archetypes from the Eloundou exposure scores\cite{eloundou2023gpts} set sector efficiency on the supply side; sector output elasticity from the energy-economics literature\cite{dimitropoulos2018rebound,saunders2013rebound,brockway2021rebound,gillingham2016rebound} modulates rebound on the demand side. We propagate seven literature-anchored parameters through Monte Carlo sampling to bound the result distributionally. Eloundou and the Anthropic Economic Index\cite{anthropic_labor_2026} serve as complementary upper-bound and current-state proxies. This study delivers the first occupation-grounded sectoral framework for adoption-side AI energy exposure. Under primary Approach~A, the US theoretical envelope is 12.1~Q and current observed deployment is $\sim 1.4$~Q (Section~2). Approach~B yields 8.8~Q and $\sim 0.4$~Q respectively; the gap localises to commercial knowledge work and supports the same directional conclusion. The largest envelope sits in industry (4.5~Q), but the largest fractional change is in transport ($+11\%$ of the 10.2~Q sector baseline, vs $+6.6\%$ in industry). We then apply the framework to all 50 US states and extend it cross-nationally to the United Kingdom.

% ═════════════════════════════════════════════════════════════════════════════
% SECTION 2: Exposure envelope
% ═════════════════════════════════════════════════════════════════════════════
\section{AI exposure across end-use energy sectors}

% P1: WHY occupations are the inferential bridge; data and SOC primer
Worker-level AI exposure can be inferred through the task profile that defines each occupation. The Standard Occupational Classification (SOC) organises $\sim$870 detailed occupations into 23 two-digit major groups (e.g., Production), 4-digit broad occupations and 6-digit detailed codes. SOC is maintained by the US Bureau of Labor Statistics and matches the BLS OEWS occupation-by-state employment data; the US Commercial Buildings Energy Consumption Survey (CBECS), Manufacturing Energy Consumption Survey (MECS) and Bureau of Transportation Statistics (BTS) provide sector-level energy baselines that we link to SOC through the crosswalk introduced here. O*NET pairs each detailed occupation with a standardised task set, providing the task-level granularity that AI exposure scoring requires. We draw on two complementary task-exposure datasets. Eloundou et al.\cite{eloundou2023gpts} score the share of each occupation's tasks that a frontier AI model could automate or augment (923 occupations; \emph{theoretical exposure} $\gamma$). The Anthropic Economic Index\cite{anthropic_labor_2026} measures the share of an occupation's tasks where Claude is used in work-related settings (756 occupations; an \emph{observed-use proxy}, denoted \emph{observed exposure} for shorthand). Merging on six-digit SOC codes yields 733 matched occupations whose scores correlate (SI~2). The theoretical metric defines the envelope's scope; the observed metric anchors it in current deployment.

\begin{figure}
    \centering
    \includegraphics[width=\textwidth]{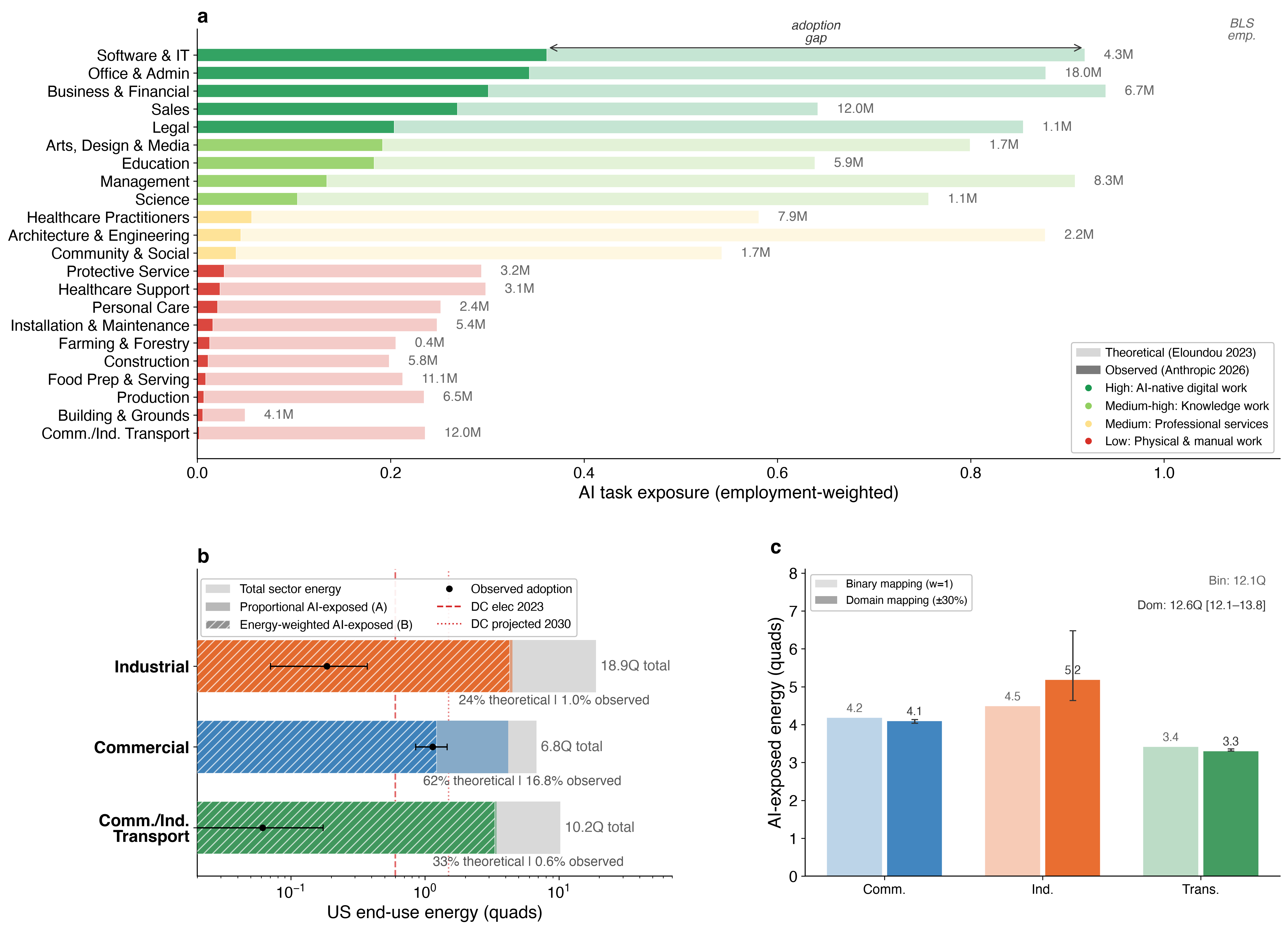}
    \caption{\textbf{US adoption-side energy exposure envelope.} \textbf{(a)} SOC major groups ranked by observed AI exposure (dark bars) and theoretical ceiling (light bars); employment annotated at right. \textbf{(b)} Sector-level envelope. Grey bars: total sector energy; solid coloured bars: economics-proportional AI-exposed portion (Approach~A); hatched bars: energy-relevance-weighted AI-exposed portion (Approach~B; SI~3); black markers: observed-adoption proxy (95\% composition-sensitivity interval from occupation bootstrap; Methods). Dashed and dotted red lines: current and projected US data-centre electricity. \textbf{(c)} Crosswalk-weighting sensitivity within the three-tier baseline (2-digit major-group defaults, 4-digit prefix overrides, named-sector 6-digit cell overrides). Light bars: binary assignment ($w = 1$); dark bars: domain-weighted with $\pm 30$\% uncertainty bars. Totals annotated as ``Bin: $X$~Q'' (binary-mapping aggregate) and ``Dom: $X$~Q [low--high]'' (domain-mapping aggregate with bracketed $\pm 30$\% range). The complementary crosswalk-resolution sensitivity, which reassigns three high-employment ambiguous SOC 53-7 cells, is reported quantitatively in Supplementary Table~S8 and discussed in the sensitivity paragraph of Section~2.}
    \label{fig:envelope}
\end{figure}

% P2: Mechanism evidence — where AI tasks are already entering energy-consuming workflows
Adoption concentrates where tasks are information-intensive and weakly coupled to physical capital. It is already extending into operational roles within energy-consuming sectors (Figure~\ref{fig:envelope}a). Software, IT and office occupations lead both the theoretical ceiling and current observed adoption, yet operate well below their theoretical potential. Headroom for further diffusion remains within the predominantly commercial-building roles they occupy. Physical and manual occupations show near-zero aggregate adoption, reflecting capability constraints on dexterity and on-site coordination\cite{metr_limitations_2026} alongside institutional barriers (capital replacement cycles, safety regulation, legacy-system integration). Within energy-consuming sectors, information-adjacent operational roles are already adopting at material rates: machine-tool programmers at 12\%, gas plant operators at 7\%, construction inspectors at 5\%\cite{anthropic_labor_2026}. This gradient implies a temporal ordering of energy-relevant effects. Effects materialise first in commercial knowledge work, then in information-intensive industrial and transport roles, and only later in physical-process occupations where rebound risk is highest. Converting this adoption signal into an energy-exposure measurement requires a transparent mapping from occupations to the sectors whose energy they influence.

% P3: HOW — crosswalk (technical justification) + bracket setup + Approach A
To assess the impact of AI usage by occupations on energy consumption, each occupation is assigned to the energy sector representing the dominant physical setting of its work. We use a three-tier crosswalk: the 2-digit SOC major-group default resolves 17 of 22 in-scope groups directly, 4-digit prefix overrides handle four of the five straddling groups (construction, maintenance, production, transportation), and 6-digit cell overrides resolve the management straddle and cells where a detailed-cell title names a single sector that contradicts its 4-digit prefix default (e.g., 47-4051 highway-maintenance workers $\to$ Industrial); see Methods and SI~4 for the qualitative justification of the three-tier design and the rule set. Two complementary approaches are used to anchor the envelope on theoretically independent grounds. \emph{Approach~A (economics-proportional)} multiplies sector energy by employment-weighted exposure, treating all employment within a sector as proportionally associated with its energy use. This matches the sectoral energy--GDP scaling implicit in standard input-output accounting\cite{stern2011energy,miller2009io} and produces an envelope directly comparable to CBECS, MECS and BTS sectoral baselines.

% P4: Approach B — energy-relevance tiers (with examples) + theoretical complementarity argument
\emph{Approach~B (energy-relevance weighted)} draws on bottom-up, workplace energy engineering: energy is proportional to each task's physical proximity to energy-consuming work, implemented through three task-energy-relevance tiers. Tasks with \textit{direct} energy relevance ($\omega = 1.0$) operate or control energy-consuming equipment: manufacturing process operators, freight drivers, HVAC technicians. Tasks with \textit{indirect} energy relevance ($\omega = 0.5$) shape energy decisions without operating equipment: facility managers, process engineers, healthcare practitioners running 24/7 facilities. Tasks with \textit{weak} energy relevance ($\omega = 0.15$) consume building energy only via occupancy and computing: office workers, financial analysts, legal and educational professionals. The 0.15 floor reflects the survey-derived share of commercial building energy attributable to computing, office equipment and occupant HVAC/lighting (SI~3). 

Approaches A and B are individually plausible but limited: Approach~A misses within-sector heterogeneity, while Approach~B carries uncertainty in the energy-relevance assignments for individual occupations. Nevertheless, approaches~A and~B together provide a complementary view of the envelope: Approach~A draws on top-down, macro-economic input-output accounting: energy is proportional to employment-weighted activity within a sector; Approach~B draws on bottom-up, workplace energy engineering: energy is proportional to each task's physical proximity to energy-consuming work. Thus, their independence makes convergence informative (the envelope is robust to the attribution choice) and divergence diagnostic (the choice matters where weak-energy-relevance work dominates and the bracket is widest).

% P5: WHAT we measure — theoretical envelope + observed envelope + convergence pattern
Using Approach~A, the theoretical envelope is 12.1~Q (Figure~\ref{fig:envelope}b): 4.2~Q commercial (62\% theoretical exposure), 4.5~Q industrial (24\%) and 3.4~Q commercial/industrial transport (33\%) (the per-sector electricity / natural-gas / petroleum / coal / other breakdown and emissions-factor implications are reported in Supplementary Table~S11). The observed-adoption envelope under Approach~A is 1.39~Q, with commercial carrying 1.14~Q and smaller contributions from industry (0.19~Q) and transport (0.06~Q). This observed envelope is already comparable in magnitude to current US data-centre electricity ($\sim$0.6~Q in 2023\cite{shehabi2024dc}) and to the IEA 2030 projection ($\sim$1.5~Q\cite{iea_energy_ai_2025}), without yet approaching the theoretical ceiling; as a scope of exposed energy rather than a consumption figure, it bounds what is at stake, while the like-for-like comparison against data-centre demand---the induced net change---is developed in Section~3. Approach~B applies an energy-relevance discount to each occupation's contribution and is reported as a complementary lens: it yields a smaller theoretical envelope of 8.8~Q and an observed envelope of 0.45~Q. The two approaches converge in industry (4.3~Q, 95\% of A) and transport (3.3~Q, 97\%), where direct-energy-relevance workers dominate; they diverge in commercial (1.2~Q, 29\%), where weak-energy-relevance knowledge workers dominate and the attribution choice carries its largest weight. Approach~B therefore supports the directional conclusion of Approach~A and localises the attribution-choice sensitivity to commercial knowledge work.

% P6: Robustness — three orthogonal sensitivity dimensions, scope statement, bridge to Section 3
The envelope is further stress-tested along three orthogonal axes, each probing a distinct judgement built into the framework: crosswalk weighting (analyst judgement on occupations that straddle two sectors; SI~4), crosswalk resolution (routing of ambiguous SOC~53-7 cells, $\sim 2.9$~million workers combined; SI~4), and the attribution anchor (Approach~A versus the energy-relevance-weighted Approach~B and the conservative-Eloundou variant; SI~2 and SI~3). Across binary and domain-weighted variants with $\pm 30$\% perturbation, the total envelope holds within $\pm 15$\% (Figure~\ref{fig:envelope}c) with sensitivity concentrating in Industrial, driven by 7.2~million SOC~53 material movers; rerouting the ambiguous SOC~53-7 cells shifts the joint-MC aggregate $\Delta E$ by 3.4\% (Supplementary Table~S8); industry and transport differ by less than 5\% under Approach~B with the attribution-choice sensitivity localised to commercial, and the conservative-Eloundou variant ($\beta = E_1 + 0.5\,E_2$) reduces the envelope by approximately one-third. All variants preserve per-sector signs, the 90\% MC range and sectoral ordering. The envelope is therefore a scope bound robust to crosswalk choice, attribution anchor and exposure-construct choice, rather than a prediction of future energy outcomes. Identical envelope sizes can produce opposite sectoral outcomes depending on how AI enters each sector's task mix and how each sector's output responds; Section~3 decomposes that dependence into a joint supply--demand model.

% ═════════════════════════════════════════════════════════════════════════════
% SECTION 3: Supply-demand decomposition of net energy outcomes (EUR-002 rewrite)
% ═════════════════════════════════════════════════════════════════════════════
\section{Efficiency, rebound and net energy outcomes}

% P1: Framework, equation, supply-demand setup
The envelope sets which energy is at stake; whether it rises or falls depends on three mechanisms. \emph{Efficiency} ($\eta$) reduces energy per unit of output when AI substitutes for or augments tasks. \emph{Rebound} ($\varepsilon$) stimulates activity within existing services when cost savings lower unit cost, so post-adoption activity equals $(1+\varepsilon)$ times pre-adoption. \emph{Enablement} ($E_{\text{new}}$) creates new energy-consuming services with no pre-AI baseline. The net energy change per sector can be estimated as:
\begin{equation}
\Delta E = \underbrace{\alpha\, E\big[(1-\eta)(1+\varepsilon)-1\big]}_{\text{efficiency--rebound on existing energy}} + \underbrace{E_{\text{new}}}_{\text{additive new demand}},
\label{eq:delta}
\end{equation}
where $\alpha \in [0,1]$ is adoption depth. Setting the first term to zero yields breakeven $\varepsilon^\ast = \eta/(1-\eta)$, separating savings from increases on existing energy; any $E_{\text{new}} > 0$ shifts the threshold toward increase. We decompose $\eta$ and $\varepsilon$ into a task-supply component (what AI can do to which tasks) and a demand-response component (how the output market responds to the cost reduction). Supply draws on the Eloundou archetype decomposition (Section~2); demand draws on the energy-economics rebound literature\cite{brockway2021rebound,saunders2013rebound,gillingham2016rebound,dimitropoulos2018rebound,wadud2016autonomous,schaller2021ride,nadel2012rebound}.

% P2: Supply side — task archetypes and sector shares (mechanism)
On the supply side, we distinguish two deployment archetypes: tasks AI performs outright (replacement) and tasks where AI assists the worker (augmentation). Eloundou et al.\cite{eloundou2023gpts} label a task E1 when a language model alone can halve its completion time and E2 when additional software built on the model is required; we read E1 as a proxy for replaceable tasks (no complementary integration is needed) and E2 as a proxy for augmentation. This mapping is an interpretive assumption of this paper rather than Eloundou et al.'s construct; the sector signs do not rest on it, because the sign thresholds derived below hold for any archetype mix (SI~6). Section~2 introduced two exposure metrics that Eloundou et al.\cite{eloundou2023gpts} report at occupation level: the theoretical exposure $\gamma = E_1 + E_2$, which counts both archetypes equally, and the conservative variant $\beta = E_1 + 0.5 E_2$, which down-weights augmentation. Reading both scalars off the published occupation table for the 923 covered occupations, the per-occupation archetype counts are recovered algebraically as $E_1 = 2\beta - \gamma$ and $E_2 = 2(\gamma - \beta)$. Aggregating these per-occupation $E_1$ and $E_2$ across occupations within each sector by 2021 BLS employment weights\cite{bls_matrix_2025} yields the sector-level archetype shares $(s_1, s_2) = (25\%, 75\%)$ for Commercial, $(33\%, 67\%)$ for Industrial and $(46\%, 54\%)$ for commercial/industrial Transport. Commercial is dominated by knowledge work where AI augments rather than replaces throughput. Industrial sits in the middle: full-substitution machine-operator and production-line tasks balance against augmentation-mode engineering, maintenance and supervisory roles. Transport's high E1 share reflects drivers, dispatchers and freight handlers mapping cleanly to full substitution, with a smaller augmentation tail in routing and scheduling.

% P3: Supply side — efficiency calibration (η_full, η_aug) with broad evidence
Sector efficiency follows as the employment-weighted (Approach A) mixture $\eta_s = s_1\,\eta_{\text{full}} + s_2\,\eta_{\text{aug}}$. The full-substitution efficiency $\eta_{\text{full}} \in [10\%, 55\%]$ spans cognitive and physical settings. Cognitive anchors include legal review, radiology and customer triage at 30--60\% per-task time reduction. Physical anchors include AI building-management at 8--40\% energy savings\cite{evans2016deepmind,ding2024buildings}, industrial AI process-control pilots at 10--40\%\cite{walther2021mlmfg} and autonomous freight\cite{wadud2016autonomous}. The augmentation efficiency $\eta_{\text{aug}} \in [3\%, 15\%]$ spans cognitive augmentation\cite{peng2023impact,noy2023experimental}, AI-augmented manufacturing\cite{berner2022germanmfg} and AI-augmented routing in trucking (UPS ORION, 8--10\%\cite{holland2017orion}). The augmentation range falls below the time-saving headlines from cognitive studies because building energy is largely persistent. HVAC, lighting and shared infrastructure consume at roughly fixed rates during occupied hours. A worker finishing a task faster frees worker time but not the building services around them, unless office space or occupied hours are also reduced. The $\eta_{\text{aug}}$ range therefore rests primarily on direct energy measurements that bypass the time-to-energy translation; cognitive time-saving studies enter as a soft upper bound rather than a literal energy-saving figure. We adopt deliberately wide bounds reflecting heterogeneity across cognitive and physical settings; refined sector-specific evidence will narrow them but is not required to establish the directional decomposition reported below (Methods; SI~6).

% P3: Demand side — sector demand-elasticity multiplier
On the demand side, how much the task-level cost reduction stimulates new activity depends on the elasticity of the sector's output market. Commercial building services saturate at comfort and working-hours thresholds: cheaper HVAC or scheduling does not materially expand demand for building space, dampening rebound\cite{gillingham2016rebound,nadel2012rebound,greening2000rebound}. Manufacturing output is more elastic: lower unit cost feeds through to larger production volumes in commoditised segments. A recent meta-analysis concludes that evidence is insufficient to assume efficiency gains reduce total industrial energy at the macro level\cite{brockway2021rebound,saunders2013rebound,bentzen2004rebound}. Freight and passenger transport are the most elastic, with long-run road-transport rebound estimated at $\sim 32\%$\cite{dimitropoulos2018rebound,small2007fuel} and autonomous-service projections of up to $60\%$ additional vehicle-miles travelled\cite{wadud2016autonomous}. Some of these anchors span ride-hail and autonomous-vehicle modes that overlap personal transport outside our commercial/freight transport baseline; the wide $\lambda_T$ range carries this mode-mixing as scenario coverage rather than mode-specific calibration. We encode this ordering through a sector demand-elasticity multiplier $\lambda_s$ applied to the task-archetype rebound base, $\varepsilon_s = \lambda_s\,(s_1\,\varepsilon_{\text{full}} + s_2\,\varepsilon_{\text{aug}})$, with $\lambda_{\text{Commercial}} \in [0.3, 0.7]$, $\lambda_{\text{Industrial}} \in [1.5, 2.5]$ and $\lambda_{\text{Transport}} \in [1.5, 2.5]$. Archetype-level rebound is calibrated at $\varepsilon_{\text{full}} \in [30\%, 100\%]$, spanning replacement-mode outcomes through backfire, and $\varepsilon_{\text{aug}} \in [2\%, 20\%]$ for augmentation-mode deployments where worker-hour bounds constrain activity but cost-driven service expansion can still occur\cite{schaller2021ride,wadud2016autonomous,saunders2013rebound,nadel2012rebound,gillingham2016rebound,azevedo2014consumer,thomas2013rebound,borenstein2015microframework} (Methods, Table~\ref{tab:joint_params}). AI-specific rebound estimates do not yet exist for these sectors. We therefore anchor priors on the cognate non-AI rebound literature with deliberately wide bounds, used here as scenario coverage rather than as a predictive distribution over AI-specific outcomes; the directional finding rests on this scenario coverage rather than any single study (SI~6).

% P4: Per-sector — mechanism and net change land together
Combining the supply and demand decompositions leaves seven parameters: two efficiency archetypes ($\eta_{\text{full}}$, $\eta_{\text{aug}}$), two rebound archetypes ($\varepsilon_{\text{full}}$, $\varepsilon_{\text{aug}}$) and three sector demand-elasticity multipliers ($\lambda_C$, $\lambda_I$, $\lambda_T$), each anchored on literature with triangular priors (SI~6). We propagate parameter uncertainty by Monte Carlo sampling ($N = 10{,}000$) at full adoption ($\alpha = 1$). Commercial sits at the low end of both axes: the lowest E1 share ($s_1 = 25\%$) and the lowest demand elasticity ($\lambda_C \sim 0.5$). Cost reductions therefore act on an inelastic, saturation-bounded demand base. AI that augments occupant productivity compounds with AI applied directly to building management on the same base, giving a net saving of $-0.22$~Q (90\% MC range $[-0.50, +0.08]$). Industrial sits in the middle of the E1 share ($s_1 = 33\%$) but at the high end of demand elasticity ($\lambda_I \sim 2$). Output expansion on energy-intensive physical processes dominates, giving a robust net increase of $+1.25$~Q (90\% MC range $[+0.49, +2.19]$). The sign is directionally clear, but the magnitude remains wide because the underlying rebound literature still disagrees by a factor of two\cite{saunders2013rebound,brockway2021rebound}. Transport combines the highest E1 share ($s_1 = 46\%$) with the highest demand elasticity ($\lambda_T \sim 2$). Driver and dispatcher tasks map cleanly to full substitution, and freight and passenger-mile demand responds sharply to cost. The net increase is $+1.12$~Q (90\% MC range $[+0.40, +2.02]$), the sector outcome that clears its breakeven most decisively of the three. This is consistent with Schaller's measured $50$--$150\%$ ride-hail vehicle-mile expansion\cite{schaller2021ride} and Wadud's autonomy projections\cite{wadud2016autonomous}, both estimated for a single transport mode (ride-hailing; autonomous personal vehicles); the adoption-side envelope extrapolates those mode-specific rebound estimates to the broader commercial-and-freight transport sector spanning road, rail, water and air.

% P6: Aggregate + macro cross-check + DC comparison + adoption-depth scaling
Sector signs are directionally robust across the literature-supported parameter space: Commercial saves in 88\% of Monte Carlo draws and Industrial and Transport each increase in over 99\%; the percentages are parameter-space robustness, not empirical probabilities. Equivalently, each sector's sign is governed by whether its demand-elasticity multiplier clears a breakeven value $\lambda^\ast_s = [\eta_s/(1-\eta_s)]/(s_1\varepsilon_{\text{full}}+s_2\varepsilon_{\text{aug}})$---or, in convention-free terms, whether AI-induced activity expansion $\varepsilon_s$ exceeds $\varepsilon^\ast_s = \eta_s/(1-\eta_s) \approx 18$--$27\%$. Because efficiency and rebound are built from the same task-archetype mix with nearly proportional calibrations, the breakeven multiplier is almost independent of that mix: it evaluates to $\lambda^\ast_s \approx 0.76$--$0.78$ ($\approx$0.8) at the Monte Carlo medians and remains within $[0.74, 1.21]$ for \emph{any} archetype mix $s_1 \in [0, 1]$---strictly between the Commercial prior ceiling ($0.70$) and the Industrial and Transport prior floor ($1.50$) (SI~6). A sector therefore increases once its demand response reaches roughly $80\%$ of the archetype-implied rebound base: a sub-proportional bar that Industrial and Transport clear by about a factor of two, while saving requires the saturation-damped response that only Commercial's demand base supports. The robustness percentages are thus prior mass on either side of this threshold---a margin rather than a point prediction---and because the threshold barely moves across sectors or mixes, the sector signs are decided on the demand side: the discriminating quantity for future end-use surveys is the demand response to AI cost reductions, not finer exposure composition. Stress-testing each parameter to its sign-flip value confirms the asymmetry: Industrial and Transport reverse only under physically implausible inputs (efficiency above $75\%$, or rebound below augmentation-mode levels) and stay majority-increase even when every prior is widened to twice the literature range, whereas Commercial reverses under modest moves and is the one genuinely borderline sector (SI~6). The aggregate median is $+2.16$~Q (90\% MC range $[+0.52, +4.12]$; under Approach~B, $+2.22$~Q with 90\% MC range $[+0.79, +3.94]$ from a parallel MC reusing the same $(\eta_s, \varepsilon_s)$ draws, SI~3). Under a strict price-channel conversion of the rebound anchors ($\varepsilon = \eta R/(1-\eta)$; SI~6), the aggregate median remains $+1.08$~Q with all sector signs intact---still roughly double current US data-centre electricity. Because the efficiency--rebound term in equation~(\ref{eq:delta}) is linear in $\alpha$ once $E_{\text{new}}$ is set to zero in the MC headline, sector signs are invariant to partial adoption; at $\alpha = 0.25$ the aggregate scales to an estimated $+0.54$~Q. As an external cross-check, Acemoglu's recent macro accounting places decadal total factor productivity gains from AI adoption at $0.5$--$2.5\%$ in an optimistic range\cite{acemoglu2024simplemacro}. Mapped to the paper's sectoral energy weights, this band places the adoption-side net change at the upper end of our distribution without further tuning, suggesting the adoption-side net energy change is aligned with predicted macro-economic productivity gains.

% P7: Enablement and substitution — qualitative, two-horizon framing
Enablement ($E_{\text{new}}$)---entirely new energy-consuming activities---and structural substitution are additive channels operating on decadal timescales beyond this paper's near-term planning horizon. Prior general-purpose technologies (electrification, information and communications technology) catalysed substantial new activity alongside substitution of existing ones, with complementary intangible-capital build-out accumulating over $20$--$30$ years\cite{autor2024new,brynjolfsson2021jcurve}. The energy footprint of AI-driven enablement differs in its sectoral asymmetry: new commercial activity tends to shift to data-centre accounting already captured compute-side, while new industrial activity (new product categories, processes) and new transport services (autonomous last-mile, on-demand logistics) consume physical energy on-site. A partial offset comes from substitution, where AI-enabled products and services displace existing ones (digital communication replacing physical mail; AI-designed lightweight components replacing heavier parts) on similar decadal timescales. We treat both effects qualitatively rather than including them in the estimated net change. The implication is a two-horizon planning view: the efficiency--rebound channel sets the near-term adoption-side trajectory, while enablement and substitution shape long-term demand composition over decades.

\begin{figure}
    \centering
    \includegraphics[width=\textwidth]{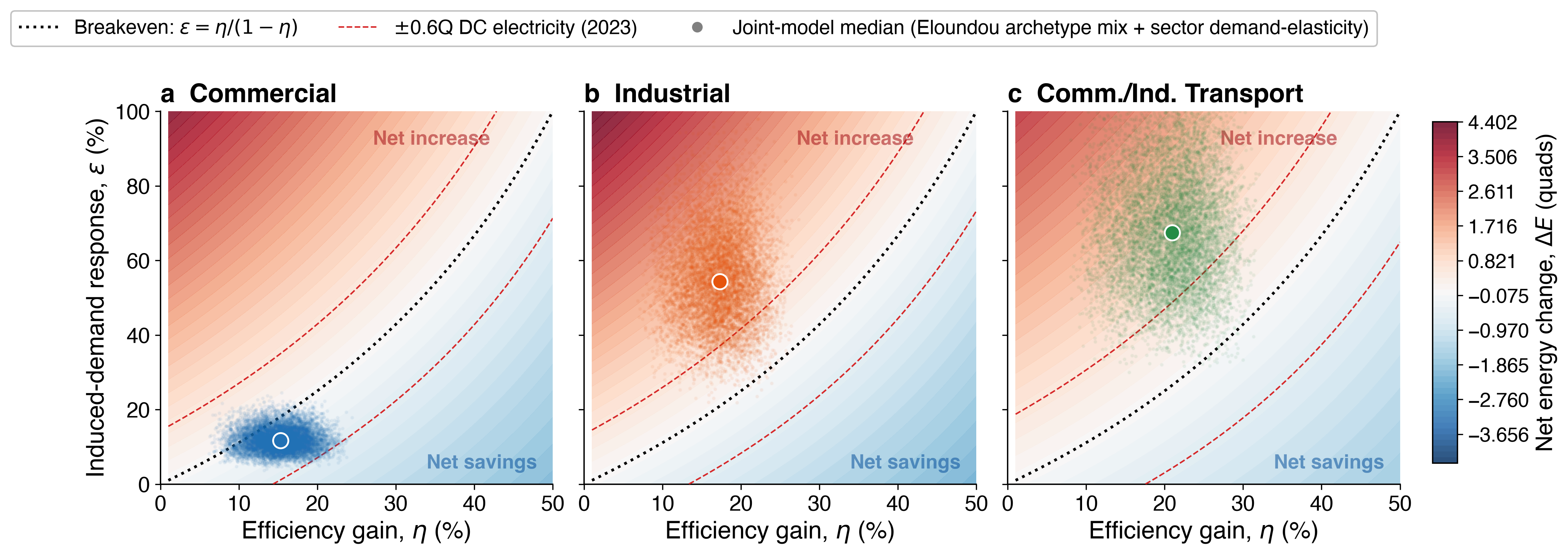}
    \caption{\textbf{Net energy change under AI adoption (joint supply--demand Monte Carlo).} Colour field shows $\Delta E$ as a function of efficiency gain $\eta$ and induced-demand response $\varepsilon$ at full adoption ($\alpha = 1$) for each sector's Approach~A exposed energy. Coloured sample clouds are 10{,}000 Monte Carlo draws of $(\eta_s, \varepsilon_s)$ from the joint supply--demand model described in Section~3 and calibrated in SI~6, with median markers (white-rimmed circles). Per-sector MC median $\Delta E$, $90\%$ Monte Carlo range (5th--95th percentile) and P(increase) are tabulated in Supplementary Table~S15. Dotted black curve: breakeven $\varepsilon^\ast = \eta/(1-\eta)$. Dashed red contours: $\pm 0.6$~Q, approximate current US data-centre electricity, 2023.}
    \label{fig:scenarios}
\end{figure}

% ═════════════════════════════════════════════════════════════════════════════
% SECTION 4: Geographic heterogeneity
% ═════════════════════════════════════════════════════════════════════════════
\section{Geographic heterogeneity}

% P1: Why regional matters + how the framework extends (two forces)
National averages can mislead regional planning when spatial heterogeneity in either energy endowment or occupational composition is large. The $+2.16$~Q national net change reported in Section~3 is an aggregate that hides where it lands. To address this, the analysis must resolve regional variation, and the task-to-energy framework supports the extension natively. Two largely independent forces then shape geographic variation: the \emph{energy endowment} of each jurisdiction (how much energy each sector consumes) and its \emph{occupational composition} (how AI-exposed its workforce is within each sector). Together they affect the envelope's size and which Section-3 sectoral scenario dominates: commercially weighted jurisdictions face a net-savings profile, while industry- and transport-weighted ones face rebound-driven increases (Methods; SI~7).

\begin{figure}
    \centering
    \includegraphics[width=\textwidth]{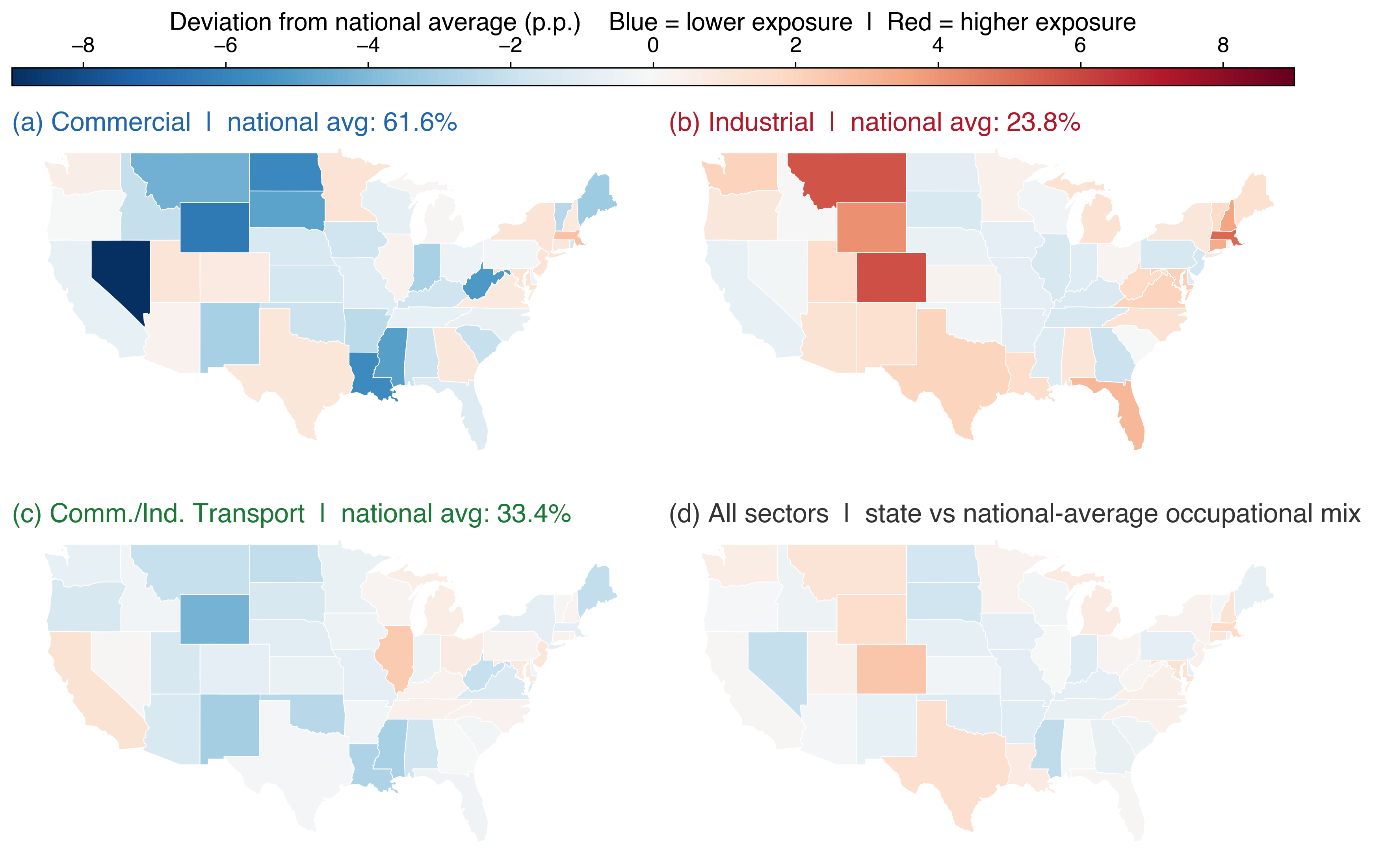}
    \caption{\textbf{US state-level deviation from national-average AI energy exposure.} Deviation (percentage points) of each state's employment-weighted exposure rate from the national average for \textbf{(a)} commercial, \textbf{(b)} industrial, \textbf{(c)} commercial/industrial transport and \textbf{(d)} all sectors combined. Red: higher-than-average exposure; blue: lower. National-average rates annotated per panel.}
    \label{fig:states}
\end{figure}

% P2: Datasets, conditions, results, projection caveat, operator pair
For the US we use EIA SEDS 2021\cite{eia_seds_2021} for sectoral energy by state and BLS OEWS 2023\cite{bls_oews_2023} for occupational employment by detailed occupation across all 50 states and the District of Columbia (Figure~\ref{fig:states}). Energy endowment varies most: industrial states such as Texas and Louisiana carry an order of magnitude more per-capita energy than services-heavy ones, shifting the envelope's centre of gravity. Occupational composition varies more subtly, with commercial-sector exposure spanning 0.53--0.70 across states and industrial exposure 0.22--0.32. The two forces compound into structurally different adoption-side exposure profiles by state. Projecting the national joint-MC sector medians ($\Delta E$ per Q exposed: $-0.054$ Commercial, $+0.276$ Industrial, $+0.323$ Transport) onto each state's per-sector exposed-energy envelope rather than re-running the joint MC at state level gives sector-decomposed and total $\Delta E$ projections per state. Industrial- and freight-heavy states sit in the rebound-risk zone where industrial and transport increases dominate small commercial savings: Texas projects $-0.02$~Q Commercial, $+0.32$~Q Industrial and $+0.13$~Q Transport for a net $+0.43$~Q; Louisiana $-0.003$, $+0.13$, $+0.03$ for net $+0.15$~Q; Indiana $-0.005$, $+0.05$, $+0.02$ for net $+0.06$~Q. Commercial-dominated states sit at the other end of the gradient. Their lighter industrial and freight base shrinks the rebound channels in absolute terms, and commercial savings are smaller still, partially offsetting rather than reversing the increase, so the net remains positive but well below the leading rebound-zone states. New York projects $-0.014$~Q Commercial, $+0.014$~Q Industrial and $+0.041$~Q Transport for a net $+0.04$~Q, comparable in magnitude to Indiana ($+0.06$~Q) despite the contrasting sector mix; Massachusetts $-0.005$, $+0.007$, $+0.016$ for net $+0.02$~Q; the District of Columbia, dominated almost entirely by commercial energy, sits closest to neutral at $\sim 0$~Q. Under this projection, planning needs differ structurally in magnitude even where the sign is the same: a Massachusetts energy office faces a small net change with commercial savings as the dominant operational shift to plan for, while a Louisiana office faces a tenfold-larger net increase concentrated in industrial facilities and freight corridors.

% P4: Generation mix overlay — framework reveals trilemma consequences (qualitative)
State exposure profiles, with the national per-sector net-change signs from Section~3 projected onto them, overlay a geographically varying generation mix. Coal and natural gas dominate the Southeast and Gulf; nuclear powers baseload in Illinois and the Carolinas; hydro supplies over 60\% of electricity in Washington and Oregon. We do not model emissions, security or cost directly, but the framework reveals where each trilemma dimension is most exposed. Industrial and transport rebound risk (Figure~\ref{fig:states}b--c) concentrates in fossil-heavy states (Texas, Louisiana, Indiana), where added load compounds carbon intensity. Above-average commercial exposure in the Northeast and Pacific (Figure~\ref{fig:states}a) pairs with cleaner generation, amplifying the gain from AI-driven efficiency. Transport rebound raises petroleum use, largely imported in the Northeast where transport exposure is also high (Figure~\ref{fig:states}c). Uneven load shifts (Figure~\ref{fig:states}d) can drive congestion pricing in transmission-constrained corridors and accelerate grid spending where adoption-side and data-centre loads compound. The framework therefore identifies where adoption-side load and generation-portfolio constraints compound, calling for regionally resolved monitoring alongside the national headline.

% P5: UK as second spatial extension — envelope transfers; MC needs UK calibration
Beyond the US, the task-to-energy framework transfers cross-nationally. We apply it to the United Kingdom via the NFER SOC~2020 $\to$ O*NET crosswalk (406 occupations matched, 365 with ONS ASHE employment data; Methods; SI~8). Combined with DESNZ sectoral energy data, this yields a UK envelope of 1.9~Q (1{,}974~PJ) out of 3{,}921~PJ national total. The exposure-rate ordering matches the US: commercial highest (1{,}016~PJ, 67\% theoretical exposure), then transport (412~PJ, 43\%) and industrial (545~PJ, 38\%). Two limitations bound further extension. UK employment data do not provide a regional dataset comparable to the ASHE-based national pipeline, so the quantitative UK case is national; an indicative ITL1 regional illustration at coarser, partially imputed 2-digit resolution (SI~8) nevertheless reproduces the US pattern, with commercially concentrated exposure in London and the South East and the largest increase-side envelope shares in Wales and the industrial North and Midlands. Although the mapping principle is jurisdiction-agnostic and AI adoption data now cover over 150 countries\cite{anthropic_geography_2025}, sub-national extension elsewhere requires regionally resolved occupational statistics that most jurisdictions do not yet publish.

% ═════════════════════════════════════════════════════════════════════════════
% SECTION 5: Discussion
% ═════════════════════════════════════════════════════════════════════════════
\section{Discussion}

% P1: Synthesis + four contributions
This study estimates the adoption-side energy exposure envelope and decomposes it into a directional net change. Net outcomes do not track envelope size: commercial holds the second-largest envelope but delivers net savings, while transport's smaller envelope drives the largest fractional increase. The framework is prior-robust: sector directional signs hold across the literature-calibrated parameter space, while the magnitude sharpens as AI-specific rebound and efficiency data accrue. The contributions are fourfold. We deliver a dual-metric exposure estimate (theoretical 12.1~Q, observed 1.39~Q under Approach~A, with Approach~B giving 8.8~Q and 0.45~Q as a complementary energy-relevance-weighted calibration). We develop a joint supply--demand decomposition yielding Monte Carlo distributions of net change with parameter-space-robust sector signs. We document a three-factor mechanism (exposure $\times$ archetype $\times$ demand elasticity) that explains sector heterogeneity. We provide an open-source framework replicable in any jurisdiction with occupational employment and sectoral energy data.

% P2: Planning implications + data infrastructure gaps
The planning implications are threefold. Energy-system models should separate compute-side load from adoption-side efficiency, rebound and enablement rather than collapsing AI into one demand category. Statistical agencies should add AI-deployment modules to end-use surveys (CBECS, MECS, freight energy accounts) and coordinate with facility-level metering already required in the EU and reported by DESNZ in the UK. Grid planners should monitor fuel displacement and load-shape changes alongside electricity, incorporating sectoral net changes alongside additive data-centre demand. Current AI energy policy in the US\cite{doe_datacenter_pathways_2024}, EU and UK\cite{iea_energy_ai_2025} focuses on data-centre supply; the envelope quantified here calls for compute-side and adoption-side planning to be coordinated rather than treated as separate problems. Tracking this heterogeneity requires data infrastructure most jurisdictions do not yet provide; four gaps tie to specific framework parameters. Energy consumption surveys should add AI-deployment indicators at facility level (sharpens $\varepsilon_{\text{full}}$ and $\lambda_s$). Occupational employment data need high resolution (4-digit SOC or finer) crossed with regional geography to support sub-national analysis (refines the archetype mix $s_1, s_2$). Task-level AI exposure data need to extend beyond US occupational classifications to avoid crosswalk noise (refines $\gamma$). And new-work measurement programmes, of the kind Autor et al.\cite{autor2024new} built for the United States 1940--2018, would directly calibrate $E_{\text{new}}$ as AI-specific evidence accrues.

% P3: Limitations + temporal-scope statement
Four limitations bound precision. The two mapping approaches are first-order estimation frameworks, not causal models\cite{eloundou2023gpts,stern2011energy}. Rebound parameters drawn from pre-AI or early-AI studies likely understate AI-driven rebound in transport and industry\cite{brockway2021rebound}; the wide priors absorb this uncertainty for sector signs but not for magnitude. The same priors also absorb transport-mode mixing, since some anchors (ride-hail, autonomous-vehicle) span personal modes excluded from the commercial/freight baseline; mode-disaggregated calibration is left to future work as AI-era mode-specific evidence accrues. Observed adoption is still concentrated in commercial knowledge work (17\% observed exposure\cite{anthropic_labor_2026}); industrial and transport effects ($<$1\% aggregate) remain largely potential, with information-adjacent roles already at 5--12\%. The exposure metric is also anchored to today's text-task capability frontier; embodied-AI advances in dexterity and on-site coordination would expand the envelope into manual and physical-process roles currently rated near-zero, broadening the industrial and transport rebound base. The mechanisms also operate at different timescales. Efficiency and rebound materialise on operational horizons and drive the $+2.16$~Q median at full adoption; the macro-consistency check in Section~3 places this within an independent decadal productivity band. Enablement and substitution emerge on decadal timescales\cite{autor2024new,brynjolfsson2021jcurve} and are weighted toward industrial and transport sectors where new physical activities consume on-site energy rather than shifting to compute load. The $+2.16$~Q headline therefore applies to the operational planning horizon; long-term composition shifts require separate analysis.

% P4: Beyond energy (brief) + closing metaphor
The framework also extends beyond energy. Adoption-side patterns coincide with non-energy distributional and productivity changes documented elsewhere\cite{anthropic_labor_2026,brynjolfsson2023generative}; these channels are not analysed here. Data-centre electricity measures the energy cost of running AI; the adoption-side envelope measures how AI-mediated task changes reshape energy flows across the broader economy. Current planning monitors the first; this paper provides a first measurement of the second.

\section*{Methods}

The analysis proceeds in three stages. Stage~1 merges occupation-level AI exposure data with US end-use energy surveys to construct the exposure envelope (Figure~\ref{fig:envelope}). Stage~2 decomposes the envelope into sector-level net energy change via a joint supply--demand model, propagated by Monte Carlo sampling (Figure~\ref{fig:scenarios}). Stage~3 extends the framework to 50 US states and the United Kingdom (Figure~\ref{fig:states}). Table~\ref{tab:methods_data} lists all primary datasets; vintage and unit harmonisation across these sources is documented in SI~9 and the industrial-scope sensitivity in SI~5.

% ── Data sources table ─────────────────────────────────────────────────────
\begin{table}
\centering
\caption{\textbf{Primary data sources.}}
\label{tab:methods_data}
\small
\begin{tabular}{@{}L{3.0cm}L{3.0cm}cL{3.8cm}@{}}
\toprule
\textbf{Dataset} & \textbf{Source} & \textbf{Year} & \textbf{Role in analysis} \\
\midrule
Theoretical exposure ($\gamma$) & Eloundou et al.\ Science\cite{eloundou2023gpts} & 2024 & Occupation-level AI exposure ceiling \\
Observed exposure & Anthropic Economic Index\cite{anthropic_labor_2026} & 2026 & Empirical adoption proxy \\
BLS employment matrix & Bureau of Labor Statistics\cite{bls_matrix_2025} & 2021 & National employment weights \\
METR benchmark (v1.1) & METR\cite{metr_horizon_2026} & 2026 & AI capability frontier (SI~1) \\
Empirical task durations & 17 published studies & 2018--2025 & Human vs AI-assisted times (SI~1) \\
CBECS & EIA\cite{eia_cbecs_2018} & 2018 & Commercial energy baseline \\
MECS & EIA\cite{eia_mecs_2018} & 2018 & Industrial energy baseline \\
BTS transport & Bureau of Transportation Statistics & 2021 & Transport energy baseline \\
Berkeley Lab DC report & Shehabi et al.\cite{shehabi2024dc} & 2024 & Data-centre reference \\
BLS OEWS (state) & Bureau of Labor Statistics\cite{bls_oews_2023} & 2023 & State-level employment \\
EIA SEDS (state) & EIA\cite{eia_seds_2021} & 2021 & State-level energy \\
NFER SOC crosswalk & Nat.\ Foundation for Ed.\ Research & 2020 & UK SOC~2020 $\to$ O*NET mapping \\
ONS ASHE & Office for National Statistics & 2024 & UK national employment by SOC \\
DESNZ TFEC & Dept for Energy Security \& Net Zero & 2021 & UK national energy by sector \\
ONS APS (regional) & Office for National Statistics & 2023 & UK regional occupational employment (SI~8 illustration) \\
DESNZ sub-national energy & Dept for Energy Security \& Net Zero & 2021 & UK ITL1 sector energy (SI~8 illustration) \\
\botrule
\end{tabular}
\end{table}

% ── 2. Exposure merge and crosswalk ──────────────────────────────────────
\subsection*{Occupation-level exposure merge}

Theoretical exposure ($\gamma = E_1 + E_2$) from Eloundou et al.\cite{eloundou2023gpts} (923 occupations) and observed exposure from the Anthropic Economic Index\cite{anthropic_labor_2026} (756 occupations) are merged on standardised six-digit SOC codes, yielding 733 matched occupations (Pearson $r = 0.59$; Spearman $\rho = 0.69$; SI~2).

\subsection*{SOC-to-energy-sector crosswalk}

The occupation-to-sector crosswalk is a contribution of this study. AI exposure is measured by occupation while energy is measured by sector: SOC tells us what workers do (computer occupations, production workers, vehicle mechanics) while EIA surveys tell us where energy is consumed (commercial buildings, industrial facilities, transport networks). The crosswalk links these two classification systems by assigning each occupation to the energy sector representing the dominant physical setting of its work. It does not claim that workers personally consume all sector energy; it uses occupation as a proxy for the operations that AI may change (office work in buildings, production work in industrial facilities, driving or maintenance work in transport systems). The Anthropic Economic Index in particular is a current-deployment proxy from observed Claude work-task interactions, not a population-wide adoption survey; we use it to anchor the theoretical envelope in current observed deployment.

The baseline mapping is a three-tier crosswalk: 2-digit major-group defaults, 4-digit prefix overrides, and 6-digit cell overrides applied in that order of precedence (6-digit $\to$ 4-digit $\to$ 2-digit). Tier~1 (2-digit defaults) resolves 17 of 22 in-scope major groups (SOC~55 Military excluded). Tier~2 (4-digit prefix overrides) handles four straddling groups (SOC~47 Construction, 49 Maintenance, 51 Production, 53 Transportation); the SOC~11 Management straddle is resolved at the 6-digit tier. Tier~3 (6-digit cell overrides) applies where a detailed-cell title names a single sector that contradicts its 4-digit prefix default (e.g., 47-4051 highway-maintenance workers and 47-4061 rail-track laying workers, whose 4-digit prefix 47-4 maps to Commercial by default; 49-3041 farm-equipment service technicians, whose 4-digit prefix 49-3 maps to Transport by default). SI~4 records the qualitative comparison against pure single-resolution alternatives (Supplementary Table~S5) and the empirical crosswalk-configuration robustness check (Supplementary Table~S8).

Three energy sectors receive occupations:
\begin{itemize}[nosep]
\item \textbf{Commercial} (6.787~Q; CBECS 2018 site energy): office-based, healthcare, education, service and building-trade occupations.
\item \textbf{Industrial} (18.906~Q; MECS 2018): farming, extraction, manufacturing, warehouse and process occupations. MECS covers manufacturing establishments; the baseline is a conservative proxy for the broader industrial sector, as non-manufacturing industrial energy (agriculture, mining) is not separately surveyed at comparable resolution.
\item \textbf{Commercial/Industrial Transport} (10.225~Q; BTS 2021): drivers, pilots, dispatchers and vehicle mechanics. This baseline covers commercial and freight modes only, excluding light-duty personal vehicles ($\sim$14.6~Q).
\end{itemize}

\subsection*{Exposure envelope computation}

Occupation-level exposure scores are aggregated to sectors using BLS 2021 employment weights\cite{bls_matrix_2025}. For each sector $s$, the employment-weighted mean exposure is:
\begin{equation}
\bar{x}_s = \frac{\sum_{i \in s} x_i \cdot w_i}{\sum_{i \in s} w_i},
\label{eq:empweight}
\end{equation}
where $x_i$ is the exposure score and $w_i$ is employment of occupation $i$. Aggregation proceeds directly from occupation-level to sector-level, bypassing the 22 SOC major groups, so that 4-digit and 6-digit overrides within a major group correctly split employment across sectors. The exposed energy in sector $s$ is:
\begin{equation}
E_s^{\text{exposed}} = E_s^{\text{total}} \times \bar{x}_s.
\label{eq:energyatstake}
\end{equation}
Two envelopes are computed: a \emph{theoretical} envelope using Eloundou $\gamma$ scores and an \emph{observed-adoption} envelope using Anthropic AEI scores. A complementary energy-relevance-weighted mapping (Approach~B) introduces occupation-specific energy-relevance weights $\omega_i$: $E_s^{\text{exposed,B}} = E_s^{\text{total}} \times \sum_{i \in s} (x_i \cdot \omega_i \cdot w_i) / \sum_{i \in s} w_i$ (SI~3). Uncertainty in the observed envelope is estimated by bootstrap resampling ($n = 10{,}000$) of occupations within each sector, recomputing the employment-weighted mean exposure per iteration. The resulting 95\% interval captures composition sensitivity (how robust the sector-level aggregate is to the specific mix of occupations and their exposure scores) rather than sampling uncertainty in the underlying scores; we therefore report it as a \emph{composition-sensitivity interval}.

\subsection*{Mapping sensitivity}

The crosswalk is stress-tested in two layers (SI~4). The published three-tier baseline is compared against a two-tier mapping (no 6-digit overrides; tiers~1 and~2 only) and a three-tier-plus-ambiguous-cell mapping that additionally reassigns three ambiguous detailed cells (53-7065 stockers, 53-7061 vehicle cleaners, 53-7081 refuse collectors); the aggregate joint-MC headline shifts by less than $0.08$~Q across these three configurations (about $3.4$\% of the headline median; the binary-mapping plus-ambiguous variant moves from $+2.165$ to $+2.092$~Q) and per-sector signs are preserved. Within the three-tier baseline and the three-tier + ambiguous-cell sensitivity variant, two weighting schemes are evaluated. Binary weighting assigns each occupation entirely to one sector; domain weighting splits ambiguous occupations between primary and secondary sectors using expert-justified weights with $\pm 30$\% perturbation bands to quantify judgement uncertainty. The eight crosswalk-by-weighting combinations (two crosswalk variants $\times$ four weighting variants) are evaluated independently. The energy-relevance-weighted mapping (Approach~B) provides an additional sensitivity dimension orthogonal to crosswalk and weighting (SI~3).

% ── 3. Joint supply-demand decomposition + Monte Carlo ──────────────────
\subsection*{Joint supply--demand decomposition and Monte Carlo propagation}

% P1: Equation + breakeven
The net energy change in each sector is
\[
\Delta E_s = \alpha\, E_s^{\text{exposed}}\big[(1-\eta_s)(1+\varepsilon_s)-1\big] + E_{\text{new}},
\]
where $E_s^{\text{exposed}}$ is the sector's exposed energy (Eq.~\ref{eq:energyatstake}), $\alpha \in [0,1]$ is adoption depth, $\eta_s$ is the fractional efficiency gain, $\varepsilon_s$ is the induced-demand response, and $E_{\text{new}} \geq 0$ is wholly new AI-enabled demand. Breakeven for the efficiency--rebound term occurs at $\varepsilon^\ast_s = \eta_s / (1-\eta_s)$, independent of $\alpha$. Because $(1+\varepsilon_s)$ scales exposed energy only, whereas a cost-driven output expansion would also draw on the sector's unexposed energy, the estimated increases for the elastic sectors are conservative in scope.

% P2: Lineage — what the model builds from
The decomposition integrates three literatures. AI task supply draws on the Eloundou archetype split\cite{eloundou2023gpts} and the Anthropic Economic Index\cite{anthropic_labor_2026}. Efficiency--rebound mechanics build on the energy-economics rebound tradition, which since Greening et al.\cite{greening2000rebound} has documented direct, indirect and economy-wide rebound across end-use sectors\cite{nadel2012rebound,gillingham2016rebound,saunders2013rebound,brockway2021rebound}. The new step couples the task-supply layer to a sector-demand-response layer, so each sector's net outcome traces back to its task mix and the elasticity of its output market. We decompose $\eta_s = s_1\,\eta_{\text{full}} + s_2\,\eta_{\text{aug}}$ and $\varepsilon_s = \lambda_s\,(s_1\,\varepsilon_{\text{full}} + s_2\,\varepsilon_{\text{aug}})$. Archetype shares $(s_1, s_2)$ follow from the Eloundou inversion $E_1 = 2\beta - \gamma$ and $E_2 = 2(\gamma - \beta)$.

% P3: Task-level efficiency η_full / η_aug + time-to-energy
We calibrate task-archetype efficiencies against AI task studies spanning cognitive and physical settings. For full substitution we adopt $\eta_{\text{full}} \in [10\%, 55\%]$. Cognitive anchors include legal review, radiology and customer triage at 30--60\% per-task time reduction. Physical anchors include AI building-management at 8--40\% energy savings\cite{evans2016deepmind,ding2024buildings}, AI process-control pilots in manufacturing at 10--40\%\cite{walther2021mlmfg} and autonomous freight\cite{wadud2016autonomous}. For augmentation we adopt $\eta_{\text{aug}} \in [3\%, 15\%]$, anchored primarily on direct energy measurements that bypass any time-to-energy translation: UPS ORION at 8--10\% measured fuel savings\cite{holland2017orion} and firm-level AI-augmented manufacturing\cite{berner2022germanmfg}. Cognitive augmentation studies\cite{peng2023impact,noy2023experimental} enter as a soft upper bound rather than a literal energy figure, because building energy is largely persistent. A worker finishing a task faster does not free HVAC, lighting or shared infrastructure during the saved time unless office space or occupied hours are also reduced (SI~6).

% P4: Task-level rebound ε_full / ε_aug + R-vs-ε note
We calibrate task-archetype rebound as $\varepsilon_{\text{full}} \in [30\%, 100\%]$, spanning replacement-mode outcomes from typical direct rebound through backfire. Saunders' 55\% manufacturing rebound\cite{saunders2013rebound} sits in the middle of this range; Schaller's ride-hail VMT expansion\cite{schaller2021ride} sits at the upper edge. We adopt $\varepsilon_{\text{aug}} \in [2\%, 20\%]$ for augmentation-mode deployments, where worker-hour bounds constrain activity but cost-driven service expansion can still occur\cite{azevedo2014consumer,thomas2013rebound,borenstein2015microframework}. We work in activity-multiplier form: post-adoption activity equals $(1+\varepsilon)$ times pre-adoption. The cited rebound literature more often reports the direct-rebound fraction $R = \Delta E_{\text{induced}}/\Delta E_{\text{engineering}}$, related to $\varepsilon$ by $\varepsilon = \eta R/(1-\eta)$ (SI~6).

% P5: Sector demand-elasticity multiplier λ_s
The sector demand-elasticity multiplier $\lambda_s$ scales the task-archetype rebound base by how elastic the sector's output market is. Commercial building services saturate at comfort and working-hours thresholds, so cheaper HVAC or scheduling does not materially expand demand: $\lambda_C \in [0.3, 0.7]$\cite{gillingham2016rebound,nadel2012rebound,greening2000rebound}. Manufacturing output is more elastic, with cost reductions feeding through to production volumes in commoditised segments: $\lambda_I \in [1.5, 2.5]$\cite{saunders2013rebound,brockway2021rebound,bentzen2004rebound}. Freight and passenger transport are the most elastic, with long-run road-transport rebound around 32\%\cite{dimitropoulos2018rebound,small2007fuel} and autonomous-service projections of up to 60\% additional vehicle-miles travelled\cite{wadud2016autonomous,schaller2021ride}: $\lambda_T \in [1.5, 2.5]$. Only the product $\varepsilon_s = \lambda_s\,(s_1\,\varepsilon_{\text{full}} + s_2\,\varepsilon_{\text{aug}})$ enters the net-change equation; the factorisation into archetype base and demand multiplier is an evidence-organising convention that lets the measured task mix, the mode-level rebound evidence and the sector demand-behaviour evidence each enter at the level where they were measured, and the identifiable claim is $\varepsilon_s$ relative to its breakeven $\varepsilon^\ast_s$ (SI~6).

% P6: Parameter table + MC propagation
Table~\ref{tab:joint_params} summarises the seven parameters and their anchors. SI~6 records the full literature classification (cognitive vs physical, replacement vs augmentation) and per-study values. We sample each parameter independently from a triangular prior with mode at the literature-anchored central value of its calibrated range (Table~\ref{tab:joint_params}); the modes coincide with range midpoints for the symmetric priors ($\varepsilon_{\text{aug}}$, $\lambda_s$) and lie off-centre for the asymmetric ones ($\eta_{\text{full}}$, $\eta_{\text{aug}}$, $\varepsilon_{\text{full}}$; SI~6). We draw $N = 10{,}000$ Monte Carlo samples at full adoption ($\alpha = 1$). For each sample we evaluate $\eta_s$, $\varepsilon_s$ and $\Delta E_s$ using the Approach~A exposed-energy values, and report median, $90\%$ Monte Carlo range (5th--95th percentile of the MC samples) and P(increase) per sector. Figure~\ref{fig:scenarios} overlays the $(\eta_s, \varepsilon_s)$ samples on a $300 \times 300$ heat-map of $\Delta E$. SI~6 reports the adoption-depth sweep and discusses additional prior-structure checks, including uniform priors and $\eta$--$\varepsilon$ correlation, as sensitivity dimensions for future reporting. $E_{\text{new}}$ is bounded from past-GPT analogues (Section~3) and discussed qualitatively; it is absent from the Monte Carlo headline.

\begin{table}[h]
\centering
\caption{\textbf{Joint supply--demand model parameters.} Triangular priors with literature-anchored modes; modes coincide with range midpoints for the symmetric priors ($\varepsilon_{\text{aug}}$, $\lambda_s$) and lie off-centre for the asymmetric ones ($\eta_{\text{full}}$, $\eta_{\text{aug}}$, $\varepsilon_{\text{full}}$). Full per-study anchors and the archetype $\times$ sector classification in SI~6.}
\label{tab:joint_params}
\small
\begin{tabular}{@{}lccL{7.6cm}@{}}
\toprule
\textbf{Parameter} & \textbf{Range} & \textbf{Mode} & \textbf{Anchors} \\
\midrule
$\eta_{\text{full}}$ & 10--55\% & 40\% & Legal review, radiology, customer triage at 30--60\% per-task time reduction; AI building-management at 8--40\%\cite{evans2016deepmind,ding2024buildings}; manufacturing AI process control at 10--40\%\cite{walther2021mlmfg}; autonomous freight\cite{wadud2016autonomous} \\
$\eta_{\text{aug}}$ & 3--15\% & 7.5\% & Cognitive augmentation\cite{peng2023impact,noy2023experimental}; AI-augmented manufacturing\cite{berner2022germanmfg}; UPS ORION 8--10\%\cite{holland2017orion} \\
$\varepsilon_{\text{full}}$ & 30--100\% & 55\% & Saunders 55\%\cite{saunders2013rebound}; Brockway et al.\ meta\cite{brockway2021rebound}; Schaller ride-hail VMT (upper edge)\cite{schaller2021ride} \\
$\varepsilon_{\text{aug}}$ & 2--20\% & 11\% & Azevedo et al.\cite{azevedo2014consumer}; Thomas \& Azevedo\cite{thomas2013rebound}; Borenstein\cite{borenstein2015microframework} \\
$\lambda_C$ & 0.30--0.70 & 0.50 & Gillingham et al.\cite{gillingham2016rebound}; Nadel\cite{nadel2012rebound}; Greening et al.\cite{greening2000rebound} \\
$\lambda_I$ & 1.50--2.50 & 2.00 & Saunders\cite{saunders2013rebound}; Brockway et al.\cite{brockway2021rebound}; Bentzen\cite{bentzen2004rebound} \\
$\lambda_T$ & 1.50--2.50 & 2.00 & Dimitropoulos et al.\cite{dimitropoulos2018rebound}; Small \& Van Dender\cite{small2007fuel}; Wadud et al.\cite{wadud2016autonomous}; Schaller\cite{schaller2021ride} \\
\botrule
\end{tabular}
\end{table}

% ── 4. Geographic extensions ─────────────────────────────────────────────
\subsection*{US state-level analysis}

State-level employment by detailed occupation is drawn from BLS OEWS May~2023\cite{bls_oews_2023}, covering approximately 830 occupations across 50 states and the District of Columbia. State-level energy by sector is from EIA SEDS 2021\cite{eia_seds_2021}. Each occupation is assigned to a sector using the identical three-tier crosswalk as the national analysis. State-level employment-weighted exposure is:
\begin{equation}
\bar{x}_{j,s} = \frac{\sum_{i \in s} x_i \cdot w_{i,j}}{\sum_{i \in s} w_{i,j}},
\label{eq:state_exposure}
\end{equation}
where $w_{i,j}$ is OEWS employment of occupation $i$ in state $j$. SEDS reports primary energy including upstream electricity generation losses, whereas national baselines use survey-specific site energy. State SEDS values are therefore normalised so that state shares sum to the national baselines: $E_{j,s}^{\text{norm}} = E_{j,s}^{\text{SEDS}} \times E_s^{\text{baseline}} / \sum_j E_{j,s}^{\text{SEDS}}$ (SI~7). The deviation $\Delta_{j,s} = \bar{x}_{j,s} - \bar{x}_{s}^{\text{national}}$ isolates the occupational-composition effect from energy endowment. We do not re-run the joint supply--demand Monte Carlo at state level; AI-deployment data at the required state-occupational resolution are not yet available. State results therefore project the per-sector $\Delta E$ signs from the national MC (Section~3) onto state envelopes via state-specific energy mix.

\subsection*{UK cross-national portability}

UK SOC~2020 four-digit occupations are mapped to US O*NET codes via the NFER crosswalk, inheriting exposure scores and sector assignments from the US pipeline (406 occupations matched; SI~8). National energy baselines are from DESNZ total final energy consumption tables (2021), with commercial transport estimated at 40\% of total transport energy. A quantitative regional extension is not pursued because UK employment data at the required occupational resolution (4-digit by region) are not publicly available; SI~8 instead provides an indicative ITL1 regional illustration at 2-digit resolution with partially imputed regional employment, reported as deviations from the UK average only.

\subsection*{Code and data availability}

All source code, source data and intermediate outputs required to reproduce the US results, every main paper figure and the supplementary sensitivity figures are released as an open reproduction bundle at \url{https://github.com/hewei8622/AI_adoption_energy}. 

\bibliography{references}

\end{document}